\documentclass[%
reprint,
aip,
amsmath,amssymb,
aps,
]{revtex4-2}
\usepackage{dcolumn}
\usepackage{bm}
\usepackage[dvipdfmx]{graphicx}
\usepackage[dvipdfmx]{color}

\newcommand{\jgr}{J. Geophys. Res.}
\newcommand{\apjl}{Astrophys. J. Lett.}
\newcommand{\grl}{Geophys. Res. Lett.}
\newcommand{\araa}{Ann. Rev. Astron. Astrophys.}
\newcommand{\mnras}{Mon. R. Astron. Soc.}
\newcommand{\ssr}{Space Sci. Rev.}
\newcommand{\aap}{A \& A}

\begin{document}
\title{Efficiency of Nonthermal Particle Acceleration in Magnetic Reconnection} 
\author{Masahiro Hoshino}
\affiliation{Department of Earth and Planetary Science, The University of Tokyo, Tokyo 113-0033, Japan.}
\email{hoshino@eps.s.u-tokyo.ac.jp}
\begin{abstract}
The nonthermal particle acceleration during magnetic reconnection remains a fundamental topic in several astrophysical phenomena, such as solar flares, pulsar wind, magnetars, etc, for more than half a century,
and one of the unresolved questions is its efficiency. Recently, nonthermal particle acceleration mechanisms during reconnection have been extensively studied by particle-in-cell simulations, yet it is an intriguing enigma as to how the magnetic field energy is divided into thermally heated plasmas and nonthermal particles. Here we study both non-relativistic and relativistic magnetic reconnections using large-scale particle-in-cell simulation for a pair plasma, and indicate that the production of the nonthermal particle becomes efficient with increasing the plasma temperature. In the relativistic hot plasma case, we determine that the heated plasmas by reconnection can be approximated by a kappa distribution function with the kappa index of approximately $3$  or less  (equivalent to $2$  or less  for the power-law index), and the nonthermal energy density of reconnection is approximately over $95 \%$ of the total internal energy  in the downstream exhaust. 
\end{abstract}
\maketitle
\section{Introduction}
Recently, the rapid magnetic energy release by magnetic reconnection, as well as the supersonic bulk flow energy release by collisionless shock waves, is regarded as an important particle acceleration process, and reconnection has garnered significant attention in various plasma environments such as the solar flares, Earth's magnetosphere, pulsar magnetospheres, and magnetars \citep[e.g.,][]{Birn07,Zweibel09,Hoshino12,Uzdensky16,Blandford17}. In these hot and rarefied magneto-active plasmas, the magnetic energy stored in a plasma sheet with an anti-parallel magnetic field component can be efficiently converted into not only the Alfv\'{e}nic bulk flow energy, but also the energies of plasma heating and nonthermal particle acceleration by magnetic reconnection.
While the reconnection paradigm has been established as a powerful magnetic energy dissipation process in the plasma universe \cite[e.g.,][]{Coppi66,Hoh66,Schindler66,Biskamp71,Galeev76}, it is recently being recognized that the reconnection plays a much more important role than before in the astrophysical phenomena, because it can efficiently generate the nonthermal particles, and the nonthermal energy spectra observed in various astrophysical objects often indicate a hard power-law spectrum, which cannot be easily explained by the conventional acceleration processes such as the diffusive Fermi shock acceleration \citep[e.g.][]{Bell78,Blandford78}.

From the satellite observations of the solar flares and substorms in the Earth's magnetotail, it is recognized that not only the plasma heating, but also the nonthermal particle acceleration is an ubiquitous process of reconnection, and it is suggested that the reconnection process with a relativistic hot gas temperature such as gamma-ray flares in the Crab nebula can generate nonthermal particles as well. In fact, by employing particle-in-cell (PIC) simulation studies, it is discussed that a hard non-thermal energy spectrum $N(\varepsilon) \propto \varepsilon^{-s}$ with its power-law index $s \le 2$ can be formed for a relativistic reconnection using 2-dimensional PIC simulations \citep[e.g.][]{Zenitani01,Jaroschek04}. It is revealed in three-dimensional PIC simulations \citep[e.g.][]{Guo14,Sironi14,Ball18,Guo20} that such a hard power-spectrum can be formed for a large magnetization parameter $\sigma=B^2/(4 \pi \rho c^2)$, whereas the power-law slope becomes softer as the magnetization parameter $\sigma$ decreases, where $B$ and $\rho$ are the magnetic field and mass density, respectively.

While the magnetic reconnection is understood as an efficient particle accelerator in plasma universe, one of the remaining challenges is to quantitatively understand the energy conversion rate into nonthermal particles. The efficiency of nonthermal particle production against the  total  internal plasma energy during the magnetic energy dissipation is not necessarily understood in a consistent way.
\citet{Ball18} discussed the nonthermal particle efficiency in trans-relativistic reconnection using PIC simulations, and determined that the efficiency increases with increasing magnetization parameter $\sigma$. However, it is interesting to know the efficiency of the nonthermal production for a wider range of  $\sigma$ values,
from non-relativistic cold plasmas to relativistic hot plasma reconnections. It is also important to carefully discuss the spectral feature from the low-energy to high-energy nonthermal range, and to understand the energy partition between the thermal and nonthermal high energies by analyzing the reconnection energy spectrum. The energy partition between the thermal plasma and nonthermal particles is still an important open question.

To understand the energy partition between the thermal and nonthermal energies, we investigate magnetic reconnection using particle-in-cell simulations by focusing on the time and spatial evolutions of the energy spectrum. The energy spectrum obtained by the reconnection simulation, in general, comprises the thermal and non-thermal populations, and a model fitting of the energy spectrum by the combination of Maxwellian and kappa distribution functions is made as well. \citep[e.g.][]{Vasyliunas68}. We quantify the thermal plasma and nonthermal populations as the function of the initial plasma temperature, both in non-relativistic and relativistic regimes, and prove that the production of the nonthermal particle becomes efficient with an increase in plasma temperature. In the relativistic hot plasma case, we determine that the nonthermal power-law index $s$ can be close to $2$, and the nonthermal energy density of reconnection reaches over approximately $95 \%$ of the total heated plasma in the exhaust. 

\section{Overview of Simulation Study}
We study the plasma heating and particle acceleration of magnetic reconnection using two-dimensional PIC simulation \cite{Hoshino87,Hoshino20,Hoshino21} with a periodic boundary in the $x$-direction for $x=-215 \lambda$ and $x=215 \lambda$, and the conducting walls for the upper and lower boundaries at $y= \pm 215 \lambda$, where $\lambda$ is the thickness of the initial plasma sheet. The total system size for our fiducial runs is $L_x \times L_y = 430 \lambda \times 430 \lambda$ and the computational grid size is $5376 \times 5376$. The total number of particles is $8.4 \times 10^{10}$, to calculate the high-$\beta$ plasma sheet as accurately as possible.

For simplicity, we adopt the Harris solution \citep{Harris62} for the pair plasma with the same mass $m$ and temperature $T$ uniform in space, and focus on the energy partition of an idealized magnetic reconnection in a collisionless plasma system. 
The magnetic field ${\bf B} = B_x(y) {\bf e_x}$ and plasma density $N(y)$ are given by
\begin{equation}
 B_x(y) = B_0 \tanh( y/\lambda),
\end{equation}
and
\begin{equation}
N(y) = N_0 \cosh^{-2}(y/\lambda) + N_{\rm b},
\label{eq:Harris_den}
\end{equation}
respectively, where $N_{\rm b}$ is the background uniform plasma density adopted to demonstrate continuous plasma injection from the outside plasma sheet to the reconnection exhaust. The background plasma density $N_{\rm b}$ is set to be $5 \%$ of the maximum Harris plasma density $N_0$. Note that the relationships of the pressure balance $B_0^2/ 8 \pi = 2 N_0 T$ and force balance  $2 T/\lambda = |e| u_d B_0/c$ are satisfied, where $u_d$ is the drift velocity , and the initial electric field ${\bf E}$ is zero.

We study eight different cases of plasma temperature from a cold non-relativistic to hot relativistic temperature of
\begin{equation}
T/mc^2 = 10^{n/2-2} \quad {\rm with} \quad n=0,1,2,..,7.
\end{equation}
The above temperatures correspond to the magnetization parameter $\sigma = B_0^2/(8 \pi N_0 mc^2)= 2 \times 10^{n/2-2}$, if we use the number density for the value of the plasma sheet $N_0$, owing to the pressure balance between the gas pressure inside the plasma sheet and magnetic pressure outside the plasma sheet.
Note that the Alfv\'{e}n speed $v_A$ is given by $v_A =c\sqrt{\sigma/(1+\sigma)}$
in a cold plasma limit, where the density is used for the value of the central plasma sheet and the magnetic field is the value of the outside plasma sheets.  We keep the ratio of the gyro-radius $r_g$ and thickness of the plasma sheet $\lambda$ for all simulation runs, and set $r_g/\lambda=0.45$, where $r_g = mc^2 \sqrt{\gamma_{th}^2-1}/(e B_0)$, 
because a thin plasma sheet whose thickness is almost equal to the gyro-radius has been formed before the explosive growth of magnetic reconnection in the pair plasma \cite{Hoshino21} and has been observed before the onset of substorms in the Earth's magnetotail \citep[e.g.][]{Asano03,Sharma08}.
For simplicity, we have assumed $(\gamma_{th}-1)mc^2/T=1$, and then  the drift speed $u_d$ can be calculated by $u_d/c=2 (r_g/\lambda)/\sqrt{1+2 mc^2/T} = 0.9/\sqrt{1 + 2 \times 10^{2-n/2}}$ .

We add a small initial perturbation for the vector potential $\delta A_z$ to the Harris equilibrium, which is given by 
\begin{equation*}
\delta A_z(x,y) \propto \exp (-(\frac{x}{\lambda_x})^2-(\frac{y}{\lambda_y})^2 )
\cosh (\frac{y}{\lambda_y}),
\end{equation*}
where $\lambda_x = 2 \lambda$, $\lambda_y = \lambda /2$, and the initial amplitude of the reconnected magnetic field is set to be $\max(B_y/B_0)=10^{-2}$ in the neutral sheet.

\begin{figure}
\includegraphics[width=8cm]{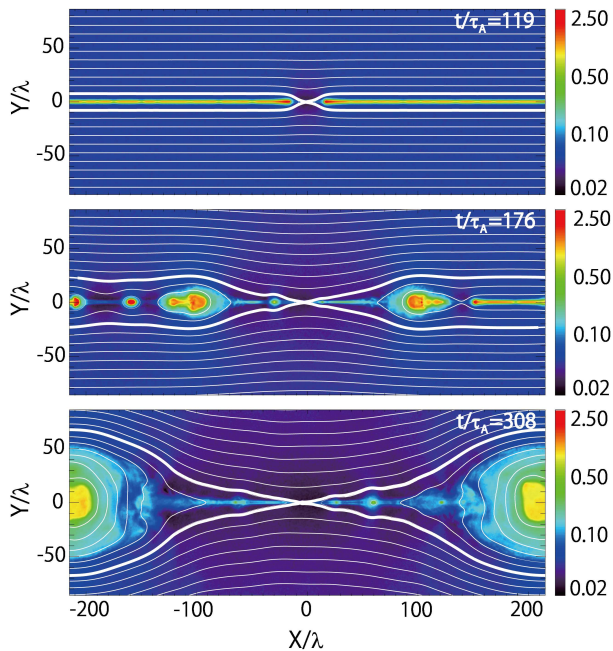}
\caption{Time evolution of the collisionless magnetic reconnection for a Harris current sheet obtained by PIC simulation
for $T/mc^2=10^{-2}$.  
The top panel ($t/\tau_A=119$) shows the early stage of reconnection, and the X-type neutral point is formed at the center of the simulation box.  The middle panel ($t/\tau_A=176$) is the nonlinear stage where several plasmoids are coalescing and forming the larger plasmoids, and the bottom panel ($t/\tau_A=308$) is the almost final stage
when two large plasmoids are just coalescing into one large plasmoid in the system.
The thick white lines indicate the magnetic field lines that pass through the X-type neutral line, i.e., the separatrix between the downstream exhaust and the upstream plasma sheet. The color bars in the right-hand side indicate that the logarithmic plasma density normalized the initial plasma sheet density.}
\label{fig:FIG1}
\end{figure}

\begin{figure*}
\includegraphics[width=18cm]{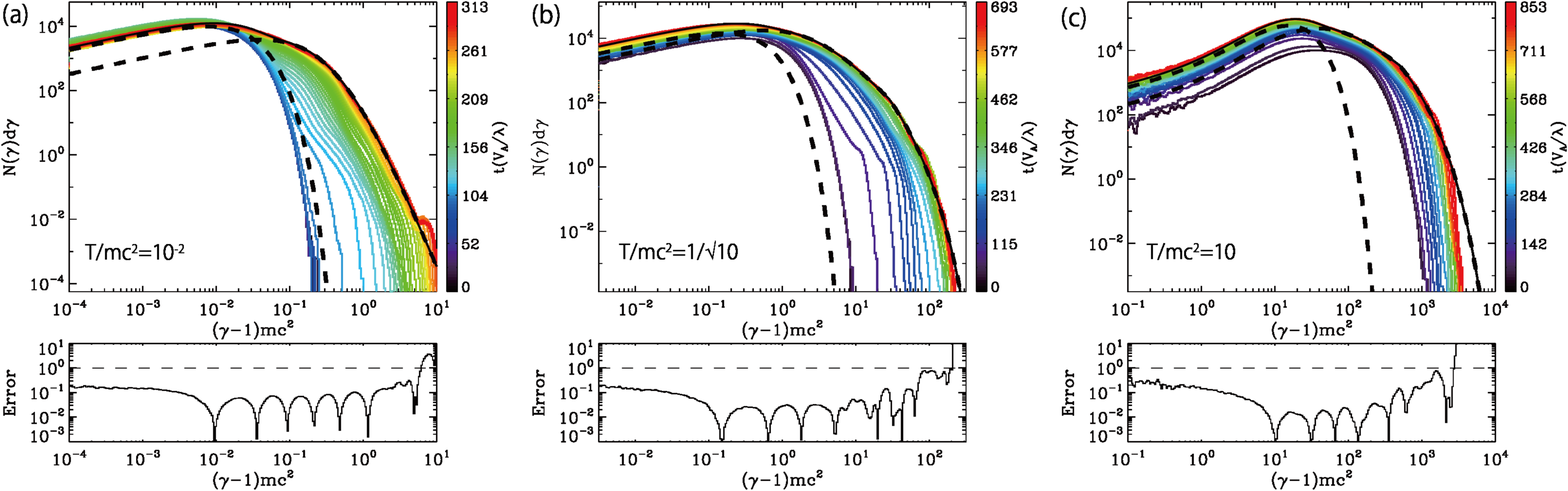}
\caption{Energy spectra for the downstream exhaust region for $T/mc^2=10^{-2}$ (left), $T/mc^2=1/\sqrt{10}$ (middle) and $T/mc^2=10$ (right), respectively. The color lines indicate the time evolution of spectra, whose time stages are indicated in the right-hand side bar. The dark blue line is the initial state, and the red one is the final stage when a large plasmoid is formed in the simulation box. The thick solid lines are the model fitting for the composed function of the Maxwellian and kappa distributions $N_{{\rm M}+\kappa}(\gamma)$. The black dashed lines in the lower energy regime represent the Maxwellian part of the fitting result $N_{\rm M}(\gamma)$, while the other dashed lines in the higher energy regime are the kappa distribution parts $N_{\kappa}(\gamma)$. The bottom panels indicate the errors of the model fitting, which show the difference between the fitting model function and simulation result. The dashed lines in the bottom plot are depicted as reference of an acceptable level of error.}
\label{fig:FIG2}
\end{figure*}

Illustrated in Figure \ref{fig:FIG1} are three-time stages of the reconnection structure obtained by the particle-in-cell simulation for a non-relativistic reconnection with the temperature of $T/mc^2=0.01$. The plasma density and magnetic field lines are indicated by the color contour and white lines, respectively. The thick white lines indicate the separatrix of the magnetic reconnection for the X-type neutral point formed around the center $x/\lambda \sim 0$, i.e., the most recently reconnected magnetic field lines. The downstream of the separatrix is occupied by the reconnection heated plasmas. These lines of the separatrix can be obtained by the contour lines of the vector potential $A_z$ that has the minimum value in the neutral line of $y=0$. A part of the simulation system is illustrated as well.

The top panel is the early time stage at $t/\tau_A=119$ when the reconnection just started at the center, where $\tau_A=\lambda/v_A$ is the Alfv\'{e}n transit time, and several small scale plasmoids have been
formed along $y=0$. The middle panel indicates the time stage when several plasmoids grow rapidly and coalesce  with  each other. As we  imposed  the initial small perturbation $\delta A_z$ at $x/\lambda=0$, the most prominent reconnection occurs at the center, but in addition to that, we can observe several other small size plasmoids embedded in the  downstream domain of the  main reconnection. The bottom panel is the approximate final stage of the active reconnection in our periodic system
 after the two large plasmoids have merged  into one large magnetic island.

\section{Model Fitting of Energy Spectrum}

As our objective is to understand the energy partition between the thermal and nonthermal populations during reconnection, we study the energy spectrum in the downstream of the separatrix lines in details. Figure \ref{fig:FIG2} illustrates the time evolutions of three energy spectra for $T/mc^2=10^{-2}$ (left), $1/\sqrt{10}$ (middle), and $10$ (right), integrated over the reconnection region sandwiched by the separatrix, i.e., the thick white lines defined in Figure \ref{fig:FIG1}. Note that if the width of the separatrix is smaller than the thickness of the plasma sheet $\lambda$, the integration is conducted for the plasma sheet with the size $\lambda$.  The horizontal axis is the particle energy $(\gamma -1)mc^2$, and the vertical axis is the number density $N(\gamma) d\gamma$, where $\gamma=1/\sqrt{1-(v/c)^2}$. The color lines indicate the time evolution of the energy spectra, whose time stages are indicated in the right-hand color bars. The blueish and reddish colors correspond to the earlier and later time stages, respectively. As we assumed a uniform plasma temperature for both the $\cosh$-type Harris density population and background populations in Equation (\ref{eq:Harris_den}), the dark blue color spectra in Figure 2 illustrate approximately the initial plasma state with the initial Maxwellian distribution functions. For the relativistic case in Figure 2c, however, one-dimensional energy spectrum is modified by the effect of a high-speed drift velocity with  $u_d/c \sim 0.82$.  

As time goes on, we can clearly observe those Maxwellian plasmas are gradually heated owing to the reconnection heating process, and the high-energy components are further accelerated and form the nonthermal population. 
During the time evolution, by looking at the energy interval of the time evolution of the energy spectra, it is determined that the energy intervals are wider in the early acceleration stage compared with those in the later stage, suggesting that the rapid energy gain happens in the early stage.

As the spectra with the red curve is approximately stable and at the final stage, we discuss the spectra behavior for the red curve by making a model function fitting in detail. It is evident that the energy spectra cannot be simply approximated by a Maxwell distribution function given by,
\begin{equation}
N_M(\gamma) \propto \gamma \sqrt{\gamma^2-1} \exp(-\frac{\gamma-1}{T_M/mc^2}),
\end{equation}
where, for simplicity, a three-dimensional Maxwell distribution function has been projected into a one-dimensional distribution in the simulation frame as the function of the particle energy $\gamma$.

To make a better model fitting of the observed distribution, we first tested a kappa distribution function given by,
\begin{equation}
N_{\kappa}(\gamma) \propto 
\gamma \sqrt{\gamma^2-1}
\left(1 + \frac{\gamma-1}{\kappa T_{\kappa}/mc^2} \right)^{-(1+\kappa)} f_{cut}(\gamma).
\label{eq:kappa_distribution}
\end{equation}
The kappa distribution comprises a thermal Maxwellian at low energies and a nonthermal population approximated by a power-law function at high energies \citep[e.g.][]{Vasyliunas68}. It is known that the observed solar wind distribution function can be well modeled by a kappa ($\kappa$) distribution function \citep[e.g][]{Vasyliunas68,Feldman75}, and the kappa modeling is widely applied for the space plasma investigation \citep[e.g][]{Livadiotis13,Lazar17}.
In our analysis, we added a high energy cutoff function $f_{cut}$ to represent a possible high energy cutoff in the simulation data \cite{Werner16,Petropoulou18}, which is given by,
\[
f_{cut}(\gamma) = \left\{
\begin{array}{@{\,}ll}
1 & \mbox{for $\gamma \le \gamma_{cut}$ } \\
\exp\left(-(\gamma -\gamma_{cut})/\gamma_{cut} \right) & \mbox{for $\gamma > \gamma_{cut}$ }
\end{array}
\right.
\]
The high energy cutoff may come from the finite time evolution of reconnection under the limited system size \cite{Petropoulou18}.
Note that the high energies of $N_{\kappa}$ can be approximated by a power-law function of $N_{\kappa} \propto \gamma^{-\kappa+1}$ with the power-law index $s=\kappa-1$ for $\gamma\gg T_{\kappa}/mc^2$.

Although the model function was not necessarily a good approximation, we determined that this kappa model function is better than the Maxwellian fitting. As the spectra for the red lines in Figure \ref{fig:FIG2} can be observed, the contributions of the cold plasma freshly transported from the outside plasma sheet is not necessarily negligible. Note that the energetic particles generated in the X-type region/magnetic diffusion region can escape along the magnetic field line, and simultaneously, the pre-heated plasma is transported from the outside plasma sheet by $E \times B$ drift motion in association with the magnetic field lines. Therefore, the separatrix region contains both the cold and hot plasmas \citep[e.g.][]{Hoshino98}.

Having the above plasma transport, we finally determined that the best-fitted model function $N_{M+\kappa}$ is a combination of the Maxwellian and kappa distribution functions, i.e.,
\begin{equation}
N_{M+\kappa}(\gamma) = N_M(\gamma) +N_{\kappa}(\gamma).
\label{eq:modelfunction}
\end{equation}

In the top three panels in Figure 2, the sold lines are the model fitting curves for $N_{M+\kappa}$ at the final time stage, while the two dashed lines are the fitting curve of $N_M$ at lower energies and $N_{\kappa}$ for higher energies, respectively. It is determined that three simulation spectra in the downstream exhaust for $T/mc^2=10^{-2}$ (left), $1/\sqrt{10}$ (middle), and $10$ (right) can be approximated by the composed model function of Maxwellian and kappa distributions $N_{{\rm M}+\kappa}(\gamma)$. We have also confirmed that five other simulation cases of $T/mc^2=10^{n/2-2}$ with $n=1,2,4,5$ and $7$ express the similar good model fitting.

The bottom three panels indicate the error of the model fitting given by,
\begin{eqnarray}
{\rm Error}(\gamma) = |N_{data}(\gamma) - N_{M+\kappa}(\gamma) |/\nonumber \\
{\rm min}(N_{data}(\gamma),N_{M+\kappa}(\gamma)),
\end{eqnarray}
where ${\rm min()}$ means a function to return the minimum element from $N_{data}$ and $N_{M+\kappa}$. If $\frac{1}{2}N_{data} < N_{M+\kappa} < 2 N_{data}$, then Error becomes less than $1$, as depicted by the dashed line in the bottom panel of Figure \ref{fig:FIG2}.
We can observe that the errors in the wide energy ranges are less than $10^{-1}$, and the model fitting of $N_{M+\kappa}$ can well describe the energy spectra obtained in the simulation runs. Note that a small bump seen around $\gamma \sim 8$ for $T/mc^2=10^{-2}$ is the contribution from the small-scale plasmoids behind the main large-scale plasmoid, but the total energy contribution is not necessarily large.
Note that the bumpy structure in the energy spectrum is not a permanent structure and disappears after the small scale plasmoid is absorbed into the large plasmoid.
For the model fitting for $T/mc^2=10$ for the right-hand panel in Figure \ref{fig:FIG2}, we can observe a large discrepancy between the simulation data and model fitting at high energies. By choosing another cutoff function given by $f_{cut}(\gamma)=\exp(-((\gamma-\gamma_{cut})/\gamma_{cut})^2)$, we obtain the better fitting result (not indicated here).

\begin{figure*}
\includegraphics[width=18cm]{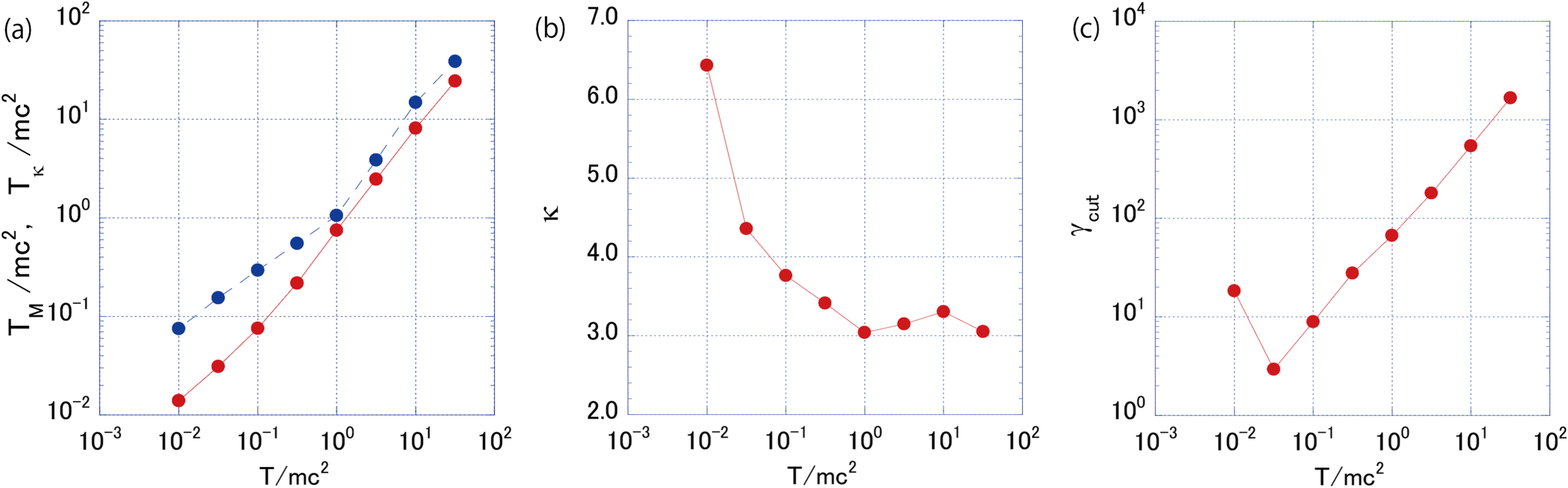}
\caption{The model fitting results for eight different initial plasma temperatures of $T/mc^2=10^{-2} \sim 10^{3/2}$. (Left) The temperatures of the Maxwellian part $T_M$ (red dots) and kappa distribution part $T_{\kappa}$ (blue dots) as the function of the initial plasma temperature. The kappa index $\kappa$ (middle) and  cutoff  energy $\gamma_{\rm cut}$ (right) as the function of the initial plasma temperature.}
\label{fig:FIG3}
\end{figure*}

We quantitatively discuss the energy partition between the thermal and nonthermal plasmas. Figure 3 illustrates the model fitting results of (a) the temperatures $T_M$ (red dots) and $T_{\kappa}$ (blue dots), (b) the $\kappa$ index with a power-law tail, and (c) the cutoff energy $\gamma_{cut}$ as the function of the initial plasma temperature $T/mc^2$, respectively. We determine that the temperature of $T_M$ remains approximately the same as the initial background temperature $T$, suggesting that the exhaust region sandwiched by the separatrix, as indicated by the thick white lines in Figure 1, still contains a part of non-heated plasmas, and
that those temperatures may represent the flux tubes just transported into the reconnection exhaust, and are not heated strongly \citep{Hoshino98}. Note that during the transport from the outside plasma sheet into the plasma sheet, the plasmas are slightly cooled down owing to the adiabatic expansion of the magnetic flux tube, because the slow mode expansion waves are emitted from the X-type reconnection region. However, the temperatures of the kappa distribution function $T_{\kappa}$ are higher than that of the Maxwellian function $T_M$. The ratio of $T_{\kappa}/T_M$ is large for the non-relativistic reconnection cases for $T/mc^2 < 1$, while those for the relativistic reconnection regime of $T/mc^2 > 1$ become less than 2.
$\kappa$ index, which is equivalent to the power-law index $s=\kappa-1$ is illustrated in Figure 3b. For the non-relativistic case of $T/mc^2=10^{-2}$, $\kappa$ indicates a large value, suggesting that the spectrum is close to a Maxwell distribution and the efficiency of the nonthermal particle production is very low, whereas for the relativistic regime of $T/mc^2 >1$, the $\kappa$ values are approximately $3$ and the nonthermal tails are well developed. Note that $\kappa=3$ can be rephrased as the power-law index of $2$.
Figure 3c is the cutoff energy $\gamma_{cut}$. Except for the case of $T/mc^2=10^{-2}$, we determine that the ratio of $\gamma_{cut}/(T_{\kappa}/mc^2)$ is approximately $50 \sim 10^2$, suggesting that the nonthermal tails are well extended above the thermal population of the kappa distribution function.

It would be interesting to mention the relationship between the gyro-radius of the accelerated particle and the size of the magnetic island.  The initial gyro-radius was set to be 0.45 $\lambda$ for all runs, and we can find in Figure \ref{fig:FIG2} that the highest energies after reconnection attained to several $10^2$ - $10^3$ times larger than the initial thermal energies.  Therefore, we can estimate that the gyro-radius of the highest energy particles is about $100 - 200 \lambda$, which is almost same as the size of the magnetic island.

It would be better to make a short comment on the model fitting error to estimate the fitting parameters.
As can be observed in Figure \ref{fig:FIG2}, the spectral profile in
the final stages of the simulation have approximately the same spectral structure in the logarithmic scale, but the fitting parameters may have approximately 10\% variations depending on the time stages. 

\section{Efficiency of Nonthermal Density and Energy Production}

\begin{figure*}
\includegraphics[width=13cm]{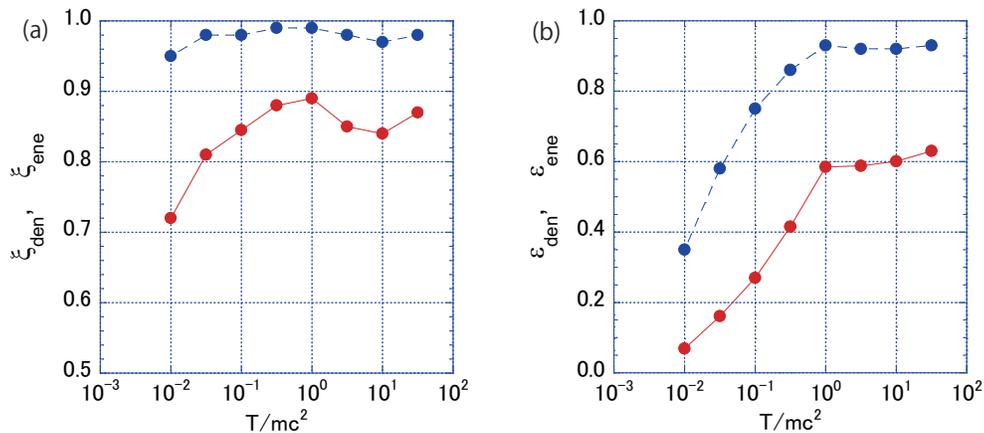}
\caption{The model fitting results for the same data set as Figure \ref{fig:FIG3}. (Left) The percentage in the number density of the kappa distribution part $\xi_{\rm den}$ (red) and percentage in the energy density $\xi_{\rm ene}$ (blue). (Right) The percentage in the number density of the nonthermal population of the kappa distribution against the total distribution $\varepsilon_{\rm den}$ (red) and percentage in the energy density of the nonthermal population $\varepsilon_{\rm ene}$ (blue).}
\label{fig:FIG4}
\end{figure*}

Based on the above model fitting, we can study the efficiency of the nonthermal particle acceleration in reconnection. In the previous section, we discovered that the energy spectrum in the exhaust region can be approximated by the combination of Maxwellian and kappa distribution functions. To quantify the efficiency of the production of the nonthermal particles, we calculate the percentage of the nonthermal population from the fitted model function.

As illustrated in Figure \ref{fig:FIG4}, the left-hand panel a is the percentage of the number density and the energy density for the population of the $\kappa$ distribution against the entire population of the energy spectrum as the function of the initial plasma temperature $T/mc^2$. The solid red and dashed blue lines are the number and energy densities defined by,
\begin{eqnarray}
  \xi_{\rm den}=
  \int_1^{\infty} N_{\kappa}(\gamma) d \gamma /
       \int_1^{\infty} N_{{\rm M}+\kappa}(\gamma) d \gamma, \nonumber
\end{eqnarray}
and
\begin{eqnarray}
  \xi_{\rm ene}=
  \int_1^{\infty} (\gamma -1) N_{\kappa}(\gamma) d \gamma /
       \int_1^{\infty} (\gamma -1) N_{{\rm M}+\kappa}(\gamma) d \gamma,  \nonumber
\end{eqnarray}
respectively. We discussed that the energy spectrum can be approximated by the combination of Maxwellian and kappa distribution functions in Figure \ref{fig:FIG2}, and the above denominator indicates the total contribution to the energy spectrum coming from both the Maxwellian and $\kappa$ portions. The percentages of the number and energy densities of the population of the kappa distribution function weakly depend on the plasma temperature $T/mc^2$, and we determine that the kappa distributions occupy more than
$95 \%$ and $70 \%$  the total energy and total number densities, respectively. The exhaust region contains the freshly injected plasma from the outside plasma sheet, but the contribution of the cold Maxwellian population is less than $5 \%$ in the total energy density. 

Figure \ref{fig:FIG4}b illustrates the ratios of the nonthermal to total number densities (solid red line), and the nonthermal energy to total energy densities (dashed blue line) as the function of the initial plasma temperature $T/mc^2$, respectively. We estimate the percentage of the nonthermal number density by the following definition of
\begin{eqnarray}
\varepsilon_{\rm den}=
\int_1^{\infty} (N_{\kappa}(\gamma)-N_{\kappa}^{\rm M}(\gamma)) d \gamma / 
       \int_1^{\infty} N_{{\rm M}+\kappa}(\gamma) d \gamma, \nonumber
\end{eqnarray}
where $N_{\kappa}^{\rm M}(\gamma)$ represents the portion of Maxwellian distribution function out of the $\kappa$ distribution function, i.e., the $\kappa$ value is replaced by $\kappa=\infty$ by keeping other fitted parameters. Therefore, the numerator of $N_{\kappa}(\gamma)-N_{\kappa}^{\rm M}$ corresponds to only the nonthermal power-law component. Note that a kappa distribution function approaches a Maxwellian distribution as $\kappa \to \infty$, as follows,
\begin{eqnarray}
\lim_{\kappa \to \infty} \left(1 + \frac{\gamma-1}{\kappa T_{\kappa}} \right)^{-(\kappa+1)} \simeq
    \exp \left( - \frac{\gamma -1}{T_{\kappa}} \right).
\end{eqnarray}
 Similarly,  the nonthermal energy density can be calculated by,
\begin{eqnarray}
  \varepsilon_{\rm ene}=
  \int_1^{\infty} (\gamma -1) (N_{\kappa}(\gamma)-N_{\kappa}^{\rm M}(\gamma)) d \gamma /\nonumber \\
       \int_1^{\infty} (\gamma -1) N_{{\rm M}+\kappa}(\gamma) d \gamma. \nonumber
\end{eqnarray}

From Figure \ref{fig:FIG4}b, it is determined that the nonthermal number and energy densities increase with an increase in the plasma temperature in the non-relativistic regime of $T/mc^2<1$, while the number and energy densities are approximately constant in the relativistic regime of $T/mc^2>1$. We determine that the relativistic reconnection can efficiently generate a lot of nonthermal particles compared with the non-relativistic reconnection. The percentage of the nonthermal number density saturates at $60 \%$, while that of the energy density is approximately $95 \%$  of the total internal energy in the downstream exhaust.  This result suggests that the energy density of the nonthermal population dominates for the relativistic reconnection with a relativistic temperature, and the thermal energy density population is negligible. The similar nonthermal efficiency has been discussed for trans-relativistic reconnection by \citet{Ball18}.

\section{Evolution of Nonthermal Plasma in Association with Magnetic Flux Tube Transport}

So far, we discussed the nonthermal energy spectra integrated all particles in the exhaust plasma sheet, which is sandwiched by the last reconnected magnetic field line; however, it would be interesting to discuss where and how those nonthermal particles are accelerated in the plasma sheet. It is known that the exhaust plasma sheet has several distinct structures responsible for particle acceleration: (i) the magnetic diffusion region with weak magnetic fields, where the fresh particles transported from the outside plasma sheet are picked up and accelerated by the reconnection electric field, (ii) the plasma sheet boundary layer between the outside and exhaust plasma sheets where the plasma gas pressure suddenly increases, (iii) the plasmoids where the high-temperature and high-density plasmas are confined by the closed magnetic flux rope, and (iv) the magnetic field pile-up region where the Alfv\'{e}nic jet from the diffusion region collides with the pre-existing plasma sheet/plasmoid etc. In association with those distinct regions, their roles on plasma heating and acceleration processes have been discussed by paying attention to the individual particle motion \citep[e.g][]{Speiser65,Hoshino01,Drake06,Hoshino98,Pritchett01,Dahlin14,Oka10}.

\begin{figure}
\includegraphics[width=8.5cm]{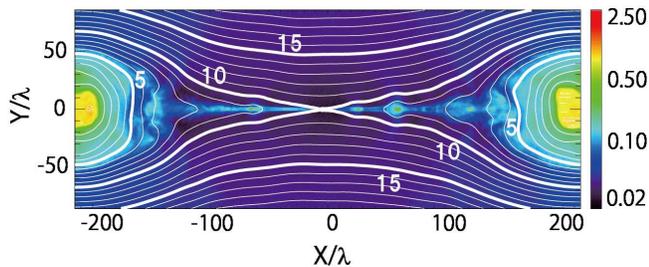}
\caption{The colour contour of the plasma density structure of magnetic reconnection for $T/mc^2=10^{-2}$. The logarithmic density contour is indicated in the right-hand side bar. The white lines indicate the magnetic field lines, i.e., the contour lines of the vector potential $A_z(x,y)$. The interval of the contour line is chosen in such a way that the downstream exhaust region is divided by ten magnetic field lines. The thick white lines are depicted with the flux tube number for every five lines, which corresponds to the spatial evolution of the energy spectrum indicated by the right-hand color box in Figure \ref{fig:FIG6}.}
\label{fig:FIG5}
\end{figure}

\begin{figure*}
\includegraphics[width=18cm]{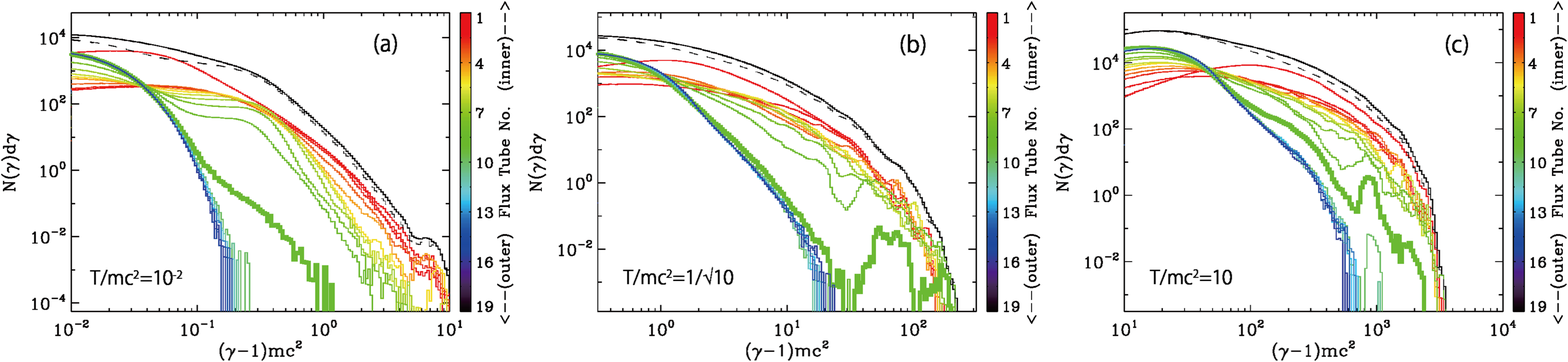}
\caption{Energy spectra for $T/mc^2=10^{-2}$ (left), $T/mc^2=1/\sqrt{10}$ (middle) and $T/mc^2=10$ (right), respectively. The same data set as Figure \ref{fig:FIG2}. The color lines correspond to the energy spectra for the 19 divided regions illustrated in Figure \ref{fig:FIG5}. The dark blue color spectrum is taken for the outermost region, while the red one is the innermost region. The right-hand side color bar indicates the region number. The solid black lines are the energy spectra for the downstream exhaust regions, which are the same spectra as illustrated in Figure \ref{fig:FIG2}.
The black dashed lines are the downstream exhaust spectra in which the contributions in the inner most flux tube are subtracted.}
\label{fig:FIG6}
\end{figure*}

The relative importance of the efficiency of nonthermal particle production in those different regions, however, has not necessarily been discussed in a  systematic  manner. To understand the relative importance, we focus on the spatial evolution of the nonthermal energy spectrum in association with the magnetic flux transportation. We divide the exhaust plasma sheet with 10 magnetic flux tubes by the equal interval of the vector potential $\Delta A_z = (A_z^{\rm max}-A_z^{\rm min})/10$, where $A_z^{\rm max}$ and $A_z^{\rm min}$ correspond to the vector potential for the last reconnected magnetic field line at the X-type neutral point, and that for the minimum bottom of the O-type point in the plasmoid, respectively. The reconnection structure divided by the 19 magnetic flux tubes for $T/mc^2=10^{-2}$ is illustrated in Figure 5, and we mark labels on the magnetic field lines from 1 – 19 from inside to outside. The last reconnected magnetic field line corresponds to the label number of 10. The underlying assumption of this analysis is that the thermal and energetic particles are transported together with the magnetic flux, because we have confirmed that the frozen-in condition is approximately satisfied \cite{Hoshino18}. 

Figure \ref{fig:FIG6} illustrates the energy spectra for those 19 divided regions labeled by the rainbow colors indicated by the right-hand side color bar. To focus on the nonthermal tail, the energy range in the horizontal axis is chosen to be three  orders  of magnitudes. The red line is the energy spectrum for the flux tube number 1, which corresponds to the innermost region almost occupied by the yellow color in Figure \ref{fig:FIG5}. The green line of the label number 10 corresponds to the flux tube just attached to the X-type neutral point, and the dark blue is that of outside the plasma sheet. Evidently, the blueish lines are basically the thermal Maxwellian spectrum with the initial plasma temperature almost maintained, thus suggesting that no plasma heating and acceleration occurs during the plasma transport from the outside plasma sheet toward the separatrix region. (Note that this statement is not necessarily correct if the evolution of the velocity distribution function is discussed in detail, because the adiabatic plasma process can be observed in association with the change of flux tube, which is initiated by the slow mode expansion waves generated around the magnetic diffusion region \cite{Hoshino18}).

However, when the magnetic flux tube is attached to the separatrix, the rapid increase in the nonthermal particle population can be observed, regardless of the initial plasma temperature. The flux numbers of 10 and 11 with the greenish colors corresponds to this region, and the flux number 11 has been highlighted by the thick green color. This behavior suggests that the initial significant particle acceleration can happen during the reconnection of two different magnetic flux tube. The flux tube number 11 is just situated outside the separatrix, but the magnetic diffusion region at the X-type neutral point has a finite width in $y$-direction, and the flux tube is contaminated by those accelerated particles around the diffusion region. 

The main energy gain in the flux tube numbers 10 and 11 comes from the meandering particles in the diffusion region \cite{Hoshino01,Pritchett01}
where the meandering particles are accelerated by the inductive electric field $E_z$.  Note that the resonant acceleration in the diffusion region can be enhanced in a relativistic plasma due to the relativistic inertia effect \cite{Zenitani01}. 
In addition to the acceleration in the diffusion region,
the beam plasma heating can occur in the plasma sheet boundary, because a part of the accelerated particles around the diffusion region may have a small pitch-angle against the local magnetic field, and those particles can escape outward along the magnetic field line. The interaction of the field-aligned beam component and cold plasma transported by $E \times B$ motion from the outside plasma sheet may lead to the plasma heating together with the generation of Alfv\'{e}nic waves \cite{Hesse18}.

After rapid energization in the magnetic diffusion region and along the separatrix magnetic field lines, those flux tubes are convected downward, and further gradual energization can be observed, as indicated by the yellowish and reddish lines. However, we determine that those energy gains are not necessarily strong, compared with those in the last reconnecting magnetic flux tube. Several bumpy structures in the spectra may arise from the dynamical evolution of plasmoids.

The thick black solid curves are the energy spectra for the downstream exhaust region integrated from the flux tube number 1 – 10, which is the same as the final stage illustrated in Figure \ref{fig:FIG2}.
By looking at the spatial evolution of the energy spectra from No.10 (flux tube of the separatrix) to No.2 (inner flux tube), we more or less observe the gradual heating and particle acceleration. However, we determine that the spectra for the innermost flux tube No. 1 has a slightly different behavior, because of the pre-existed and pre-heated plasma before the onset of reconnection. To distinguish the plasma heating between the pre-existed plasma and freshly injected particles, we indicate that the energy spectra integrated the region from the flux tube of No.2 – No.10 as the black dashed line, i.e., we excluded the innermost flux tube. As can be observed, the difference appears only around the middle energy range between the thermal Maxwellian part and nonthermal power-law part, and the kappa index becomes slightly harder as well. Regarding the case of the relativistic reconnection, we obtain the kappa index of $\kappa < 2$, and the result is roughly consistent with the previous studies \cite{Zenitani01,Jaroschek04,Guo14,Sironi14}. The efficiency of nonthermal density discussed in the previous section, however, does not change so much.

\begin{figure*}
\includegraphics[width=18cm]{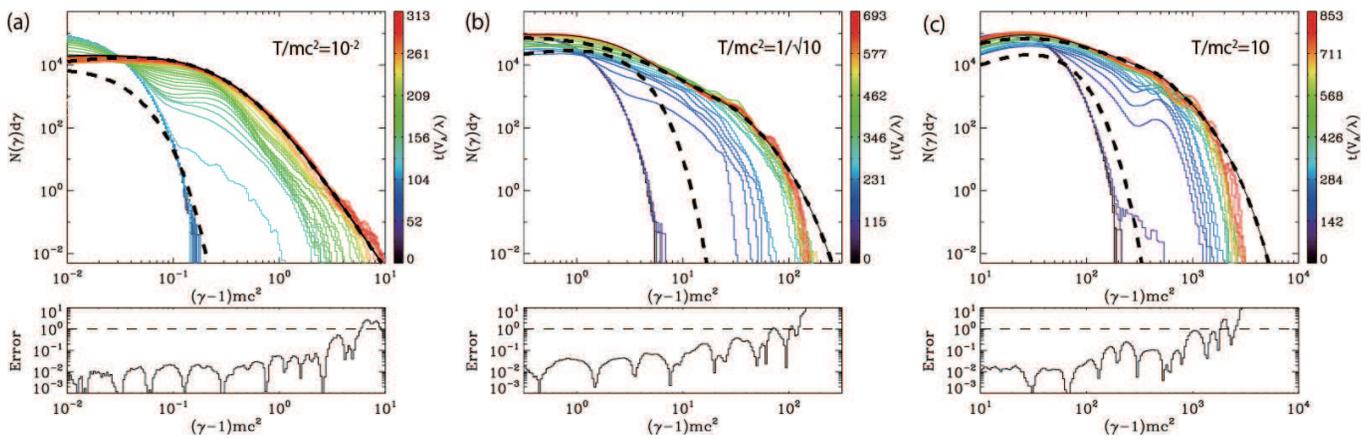}
\caption{
Time evolution of energy spectra for fixed magnetic flux tubes for $T/mc^2=10^{-2}$ (left), $T/mc^2=1/\sqrt{10}$ (middle) and $T/mc^2=10$ (right), respectively. The same data set as Figure \ref{fig:FIG2}.
The color lines indicate the time evolution of spectra, whose time stages are indicated in the right-hand side bar. The dark blue line is the initial state when the magnetic flux tube is situated in the upstream.  On the other hand, the red one is the final stage when the flux tube is transported into the downstream exhaust.  
The thick solid lines are the model fitting for the composed function of the Maxwellian and kappa distributions $N_{{\rm M}+\kappa}(\gamma)$. The black dashed lines in the lower energy regime represent the Maxwellian part of the fitting result $N_{\rm M}(\gamma)$, while the other dashed lines in the higher energy regime are the kappa distribution parts $N_{\kappa}(\gamma)$. The bottom panels indicate the errors of the model fitting in the same format as Figure \ref{fig:FIG2}.
\label{fig:FIG7}
}
\end{figure*}

Although we discussed the spatial evolution of the energy spectra in Figure \ref{fig:FIG6}, it is also interesting to mention the time evolution of the energy spectrum for a fixed magnetic flux tube, which is transported from the outside plasma sheet into the downstream exhaust region 
during reconnection.
Figure \ref{fig:FIG7} illustrates the time evolution of the energy spectra for the fix magnetic flux tubes sandwiched between $\left| y \right|/\lambda=9$ and $12$ at the initial state $t=0$.  
As could be easily expected, the behavior of the energization is basically same as Figure \ref{fig:FIG6}.  The dark blue color spectra are taken when the flux tubes are situated in the upstream, and those temperatures almost remain the initial Maxwellian.  As time goes on, we observed the rapid energization and formation of the nonthermal tail when two flux tubes just reconnect at the X-type neutral point at $t/\tau_A \sim 120$.  The timing of the attachment of two flux tubes normalized by the Alfv\'{e}n transit time $\tau_A$ was almost same for three cases.

After the attachment of the two magnetic flux tubes, those reconnected magnetic flux tubes are further transported into the downstream, and the high energy particles are gradually produced.  The red lines are the almost final stages of the magnetic reconnection.  The thick solid lines are the model fitting for the composed function of the Maxwellian and kappa distributions $N_{{\rm M}+\kappa}(\gamma)$. The black dashed lines in the lower energy regime represent the Maxwellian part of the fitting result $N_{\rm M}(\gamma)$, while the other dashed lines in the higher energy regime are the kappa distribution parts $N_{\kappa}(\gamma)$.  The fitted parameters of ($T_M/mc^2$, $T_{\kappa}/mc^2$, $\kappa$, $\gamma_{cut}$) are ($0.013$, $0.078$, $6.1$, $6.2 \times 10^2$), ($0.79$, $0.44$, $3.0$, $2.2 \times 10^1$), and ($15.$, $9.0$, $2.6$, $4.7 \times 10^2$) for $T/mc^2=10^{-2}$, $1/\sqrt{10}$, and $10$, respectively.  These fitted parameters are basically same as those obtained in Figures \ref{fig:FIG2} and \ref{fig:FIG3}.  The contributions of the Maxwellian part, however, are smaller than the kappa distribution part compared with Figure \ref{fig:FIG2}, because the magnetic flux tubes are convected into the deep downstream and the upstream cold plasma are heated.  The bottom panels indicate the errors of the model fitting in the same format as Figure \ref{fig:FIG2}. We find that the model fittings are very good for a wide energy ranges.

\section{Discussions and Summary}
During the last two decades, it has been discussed that magnetic reconnection can produce not only the hot thermal plasma, but also the high-energy supra-thermal particles. Specifically, relativistic reconnection can efficiently produce nonthermal particles, whose energy spectrum is approximated by a hard power-law function. However, how the efficiency changes between the non-relativistic and relativistic reconnection has remained an  open  question. Meanwhile, the detailed acceleration process focusing on the individual particle motion has been well discussed by several people \cite[e.g.][]{Hoshino98,Hoshino01,Drake06,Pritchett01,Dahlin14,Oka10}.

The interaction between the reconnection electric field $E_z$ and meandering motion under the anti-parallel magnetic field $B_x$ in the neutral sheet is known to play an important role as the primary acceleration process, and the accelerated particles in and around the X-type neutral line can be ejected outward by the Lorentz force because of the reconnecting magnetic field \citep{Speiser65,Pritchett01}. \citet{Zenitani01} investigated this process to the relativistic reconnection, and argued that the interaction efficiency between the reconnection electric field and meandering particle can be enhanced in the relativistic reconnection, because of the relativistic inertia effect during the Speiser motion \citep{Speiser65}.

In addition to the particle acceleration in and around the magnetic diffusion region, several other possible mechanisms have been proposed. The collision between the Alfv\'{e}nic jet ejected from the magnetic diffusion region and magnetic field pile-up region ahead of the plasmoid can provide the betatron-like acceleration owing to the gradient-B drift motion \citep[e.g][]{Hoshino01}. \citet{Drake06} proposed that the particles trapped by the plasmoid can be energized by a Fermi process, which is the curvature drift motion in association of the time evolution of the plasmoid. \citet{Drake06,Dahlin14} investigated the above mechanisms responsible for energy gain using the guiding center formalism. \citet{Oka10} argued that the particles can be accelerated during coalescence of two plasmoids with the similar process that occurs in the magnetic diffusion region, but with a stronger inductive electric field in association with the dynamic coalescence.

We did not analyze such detail acceleration mechanisms in this study, rather focused on the evolution of energy spectrum during the plasma transport in association with the magnetic flux tube. We not only observed the primary particle acceleration around the magnetic diffusion region, but also the additional acceleration during the plasma transport in the exhaust. These heating and acceleration are basically consistent with the previous studies. In addition, we determined that the energy spectra obtained by PIC simulations can be well approximated by a kappa distribution function for the high energy population. By examining the model fitting results, we determined that the efficiency of the nonthermal particle acceleration increases with increasing plasma temperature, and the nonthermal energy density reaches over $95 \%$ of the total internal energy in the relativistic temperature  of the exhaust,  whereas the nonthermal number density remains approximately $40 \% \sim 60 \%$ for the total number density
 in the exhaust. 

Our simulation study in a two-dimensional system with the anti-parallel magnetic field topology is only a  first  step toward understanding the energy partition between thermal and nonthermal plasmas.
We investigated the efficiency of the nonthermal particle production for a pair plasma, but it is important to understand the acceleration efficiency for an ion-electron plasma. Recently, several people discussed the energy partition between ion and electron, and it is revealed that ions are preferentially heated, compared to electrons during magnetic reconnection by PIC simulations \cite{Shay14,Haggerty15,Rowan17}. It has been discussed that the ratio of the ion heating to the electron $T_i/T_e$ can be approximated by $(m_i/m_e)^{1/4}$ for the magnetic reconnection without guide field, where $m_i$ and $m_e$ are the ion and electron mass, respectively \cite{Hoshino18}, whereas the electrons receive larger energies than the ions for guide field reconnection \cite{Rowan19}. Although ions seem to receive much thermal energies compared with electrons, suprathermal electrons seem to be efficiently generated, while the proton nonthermal spectrum seems not to be necessarily well developed in particle simulations. Significant attention has been paid to electron nonthermal particle acceleration, but the study of ion nonthermal production is limited. We are probably required to study a longer time evolution of reconnection for a large system size.

 It is important  to check the dependence of the energy spectrum on the system size of our PIC simulation.
So far, we fixed the system size with (a) $430 \lambda \times 430 \lambda$, but we studied other cases with (b) $215 \lambda \times 215 \lambda$, (c) $860 \lambda \times 215 \lambda$, and (d) $1720 \lambda \times 430 \lambda$ for $T/mc^2=0.1$. The fitting results for $T_M$, $T_{\kappa}$, $\kappa$,
 $\varepsilon_{\rm den}$ and $\varepsilon_{\rm ene}$ 
does not strongly depend on the system size, and
we found that those differences of $T_M$, $T_{\kappa}$, $\kappa$ on the system size
were less than $15 \%$. However, the cutoff energy seems to weakly depend on the system size, except for case (b). Regarding case (b), $\gamma_{\rm cut}$ was approximately 6.7, but other cases of (a),(c) and (d) were $\gamma_{\rm cut} \sim 9-12$. 

It would be important to mention the effect of the plasma temperature in the outside plasma sheet, which is denoted by $N_b$ in Equation (\ref{eq:Harris_den}). So far, we assumed that the background temperature is the same as that for the Harris plasma component $N_0$, but we can choose any temperature to keep an equilibrium state. In astrophysical settings, it is likely that the plasma temperature outside the plasma sheet is colder than the Harris plasma temperature. In fact, for the case of the Earth's magnetotail, the temperature in the lobe is known to be colder than that in the plasma sheet \cite[e.g.][]{Asano03,Sharma08}. We have studied the cold background plasma temperature $T_b/mc^2=10^{-4}$ by keeping other plasma parameters the same. We determine that the efficiency of the nonthermal energy density is almost the same as the case of the uniform background plasma temperature discussed in this study. We will discuss this behavior in a separate paper.

Our final comment is on the effect of the background density $N_b$. It is well known that the reconnection rate increases with an increase in the Alfv\'{e}n speed in the inflow region. Furthermore, the reconnection electric field $E_z$ becomes larger with a decrease in the background density, and the magnetic energy dissipation rate being converted into the thermal and nonthermal plasmas would be enhanced as well. Although the nonthermal particle production would be expected to increase with a decrease in the background plasma density, the energy partition between the thermal and nonthermal plasma production should be carefully studied.
In our preliminary study for the low background density with $N_b/N_0=1/80$, we obtained the nonthermal $\kappa$ values of $2.2$, $2.6$, $2.7$, and $4.3$ for $T/mc^2=10$, $1$, $10^{-1}$ and $10^{-2}$, respectively.  We find that these $\kappa$ values are about $30 \%$ smaller than those of $N_b/N_0=1/20$ shown in Figure \ref{fig:FIG3}b.
It would be also interesting to study the effects of background temperature and density for the generalized magnetization parameter $\sigma = B^2/(4 \pi w)$ and Alfv\'{e}n speed $v_A=c/\sqrt{1+4 \pi (e+p)/B^2}$, where $w$, $e$ and $p$ are the enthalpy density, the total energy density and the gas pressure, respectively. 
We will discuss these effects in a separate paper together with the effect of the guide field. 

\acknowledgments
This study was supported by JSPS KAKENHI, Grant Number 20K20908.
\section*{DATA AVAILABILITY}
The data support the findings of this study are available from the corresponding author upon reasonable request.
\appendix
\section{Nonthermal contribution in $\kappa$ distribution function}

\begin{figure}
\includegraphics[width=7cm]{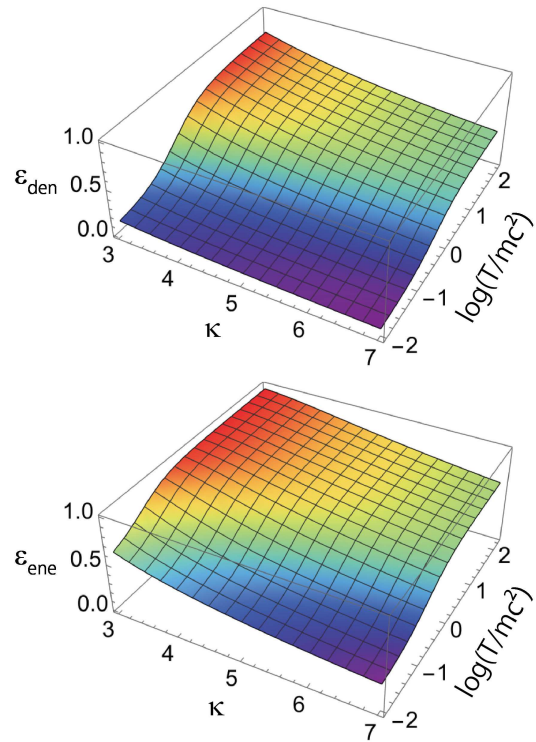}
\caption{The percentage of the nonthermal number density $\varepsilon_{\rm den}$ (top) and nonthermal energy density $\varepsilon_{\rm ene}$ (bottom) as the functions of the temperature $T/mc^2$ and kappa index $\kappa$.}
\label{fig:FIGA1}
\end{figure}

We shortly discuss the basic behavior of $\kappa$ distribution function by focusing on the thermal and non-thermal populations. As the $\kappa$ distribution function are given by two parameters of the temperature $T$ and $\kappa$ index, the nonthermal contribution can be given by two parameters. By assuming that the thermal population of the $\kappa$ distribution function is given by keeping the same temperature but by setting $\kappa=\infty$, we can easily estimate the percentages of the nonthermal number and nonthermal energy densities. The top and bottom panels in Figure \ref{fig:FIGA1} illustrate the nonthermal number density $\varepsilon_{\rm den}$ and the nonthermal energy density $\varepsilon_{\rm ene}$, respectively. The horizontal axes are the $\kappa$ index and plasma temperature normalized by the rest mass energy $T/mc^2$. We can observe that the nonthermal contributions become small for a large $\kappa$ value. For the case of a small $\kappa$ and relativistic hot plasma, we determine that the nonthermal contributions become significant.
Regarding $\kappa=3$, the nonthermal number density contribution $\varepsilon_{\rm den}$ becomes $\varepsilon_{\rm den} \rightarrow 7/9$ for $T/mc^2 \rightarrow \infty$, whereas the nonthermal energy density contribution is $\varepsilon_{\rm ene} \rightarrow 1$ for $T/mc^2 \rightarrow \infty$. For $\kappa=4$ and $T/mc^2 \rightarrow \infty$, $\varepsilon_{\rm den} \rightarrow 5/8$ and $\varepsilon_{\rm ene} \rightarrow 29/32$. 

It is also interesting to note the effect of the cutoff energy function introduced in the simulation analysis. In our simulation results, the cutoff energies were much larger than the heated plasma temperature, i.e., $\gamma_{cut}-1 \sim 10^2 \times T/mc^2$. By assuming that $\gamma_{cut}-1 = 10^2 \times T/mc^2$, we determined that the difference between the nonthermal contribution with and without the cut-off energy is less than several $\%$ for $\kappa > 3$.

\end{document}